\documentclass[12pt]{iopart}

\usepackage{graphicx}
\def\Journal#1#2#3#4{{#1} {\bf #2 }(#3) #4}

\def\NIMA{{ Nucl. Instrum. Methods} A}
\def\NPB{{ Nucl. Phys.} B}
\def\NPA{{ Nucl. Phys.} A}
\def\PLB{{ Phys. Lett.}  B}
\def\PRL{ Phys. Rev. Lett.}

\def\PRC{{ Phys. Rev.} C}
\def\PRD{{ Phys. Rev.} D}
\def\ZPC{{ Z. Phys.} C}

\def\EJPC{{ Euro. J. Phys.} C}

\begin{document}

\title{High $p_T$ Identified Particle Spectra\\
\normalsize{-- the effective color-charge factor}}
\author{Zhangbu Xu}
\address{Physics Department, Brookhaven National Laboratory, Upton, NY 11973, USA}
\ead{xzb@bnl.gov}
\begin{abstract}I will present an overview of identified particle spectra at high $p_T$ ($p_T{}^{>}_{\sim}$ 5 GeV/$c$) in both p+p collisions and AA collisions at RHIC. In p+p collisions, summary of particle ratios of K, $\eta$, $\omega$, $\rho$, $\phi$, $p$, $\bar{p}$, $\Lambda$ and heavy-flavor (open charm, $J/\Psi$) to $\pi$ at high-pt will be compiled and compared to the ratios of integrated yields. The spectra are used in $x_t$ scaling study and compared to pQCD calculations. These will help us establish particle composition in jets and the quark and gluon contributions to hadron production at high $p_T$. Similar jet chemistry has been extracted in Au+Au data in search for a quantitative measure of color charge dependence of jet energy loss. 
\end{abstract}
\section{Introduction}
\subsection{Fundamentals in QCD}
There are two fundamental questions in QCD that have motivated the relativistic heavy-ion collisions: quark confinement and symmetry breaking~\cite{lee95}. There is asymptotic freedom, where the coupling ($\alpha_s$) of the strong interaction becomes weaker at shorter distance and higher energy, while the required energy to pull color objects apart grows with distance(quark confinement). The fact that gluons carry color charge also has profound consequences: gluons can interact strongly among themselves and with quarks, which generates $>98\%$ of the masses of hadrons(symmetry breaking). SU(3) is the right group for QCD. This provides an accurate account of color-charge factor ("interaction strength") of the following three Leading Order processes: quark emitting a gluon ($\alpha_sC_F$), gluon splitting into two gluons ($\alpha_sC_A$), and gluon splitting into quark-anti-quark pair ($\alpha_sT_F$). The SU(3) predicts $C_A/C_F=9/4$ while experiments with $e^+e^-$ collisions at 91 GeV at LEP obtain a value of $2.29\pm0.06(stat.)\pm0.14(syst.)$~\cite{lepcolor} from multiple jet analyses. This provides a stringent constraint that SU(3) is the right group theory of QCD. In a four-jet event in $e^+e^-$ analyzed for the observation of color-charge factor, each jet has an average energy of 23 GeV, not very different from jet produced by p+p and A+A collisions at RHIC~\cite{starppjet}.

\subsection{Monifests of QCD properties on high-$p_T$ hadron spectra}
How would the color-charge factor manifest in heavy-ion collisions? In jet quenching scenario, an energetic quark (gluon) emits a gluon due to interaction with QGP~\cite{jetquenching}. To leading order, the diagram is the same as quark (gluon) emits a gluon in $e^+e^-$ collisions. Therefore, the energy loss is proportional to $\alpha_sC_{[A,F]}<\hat{q}>L^2$, where $C_A(C_F)$ is for quark (gluon) jet. It is obvious that if pQCD is applicable to the energy loss, the different energy loss between quark and gluon has to be $C=C_A/C_F=9/4$. Heavy quarks are a different beast (a nice feature though) since the gluon radiation is suppressed at small angle because the emission rate is inversely proportional to $((M/E)^2+\theta^2)$ (dead cone effect)~\cite{deadcone}. In $5^{<}_{\sim}p_T{}^{<}_{\sim}20$ GeV/c, the charm quarks show similar behavior as light quarks ($u,d,s$) while bottom quarks are much less suppressed due to dead cone effect. This provides excellent observables to test one of the basic ingredient of QCD: whether SU(3) in QCD is still the most relevant effective group in strongly interacting Quark-Gluon Plasma. We emphasize the {\bf "effectiveness"} since the author doesn't believe we are testing the correctness of the SU(3) for QCD. The anolog is that there are many effective theories for QED in the condense matter even though the QED is the correct theory for electromagnetic interaction. In reality, it is not as simple as this, there are geometry and pathlength fluctuation in jet quenching. The WHDG model has taken this into account, and shows that jet quenching at parton level still proportional to this color-charge factor~\cite{WHDG,xnwang,renk}. In addition, an energetic quark (gluon) can have Compton-like scattering with the partons in QGP, providing a flavor-changed ($q\rightarrow q$ or $g\rightarrow q$) leading parton~\cite{liuconversion} (details in discussion section~\ref{discussion}). However, due to confinement, we are not able to directly observe quark or gluon in an experiment. Energetic partons fragment into cluster of hadrons (jet). Jet reconstruction therefore provides the closest observable to single energetic partons. Indeed, this is the basic tool for the measurements of color-charge factor and for many observations and discoveries related to QCD or beyond QCD~\cite{lepcolor,CDF,starppjet} in high-energy $e^+e^-$ and hadron collisions. 

The production of hadrons from jet can be separated into three distinct terms in a naive picture: parton distribution function(PDF), parton interaction cross-section at $2\rightarrow2$($gg\rightarrow gg$,$qq\rightarrow qq$,$qg\rightarrow qg$, $gg\rightarrow q\bar{q}$, $q\bar{q}\rightarrow gg$, and $q\bar{q}\rightarrow q\bar{q}$), and parton fragmentation function~\cite{jetquenching}. In principle, PDF is provided by DIS e+p collisions, $2\rightarrow2$ amplitude is provided by pQCD theory, and FF is from $e^+e^-$. If we assume: those three terms can be factorized, PDF and FF are measured with sufficient accuracy, and FF has universality, we can predict what hadron spectra at high $p_T$ should be in p+p collisions at RHIC. Are these ingredients sufficient for p+p collisions? 
how will measurements in p+p collisions provide additional information for our understanding of QCD and for model development? What will be modified in A+A collisions: PDF, FF, $\alpha_s$ and the effective color-charge factor? 

In heavy-ion collisions at RHIC, the energy is sufficient to produce a well-defined jet at initial stage as in p+p collisions~\cite{starppjet}. However, the soft processes, which are necessary for QGP creation, produce overwhelming background and prevent a meaningful full-jet reconstruction to date. Instead, we rely on leading hadrons to identify and study the effect of jet quenching. This inevitably requires a detailed understanding of the fragmentation function (FF) of the partons to hadrons. We know from detailed measurements in $e^+e^-$ at LEP and SLAC that gluon and quark jets have distinct features when fragmenting. In general, gluons produce more soft particles and more leading baryons than quarks do. Flavor separated fragmentation functions from quark and gluon into identified pions have been provided by $e^+e^-$ data, and have been tested extensively at hadron colliders~\cite{CDF,phenixpi0,starPIDpapers}. 
\begin{figure}
{\includegraphics[scale=0.6]{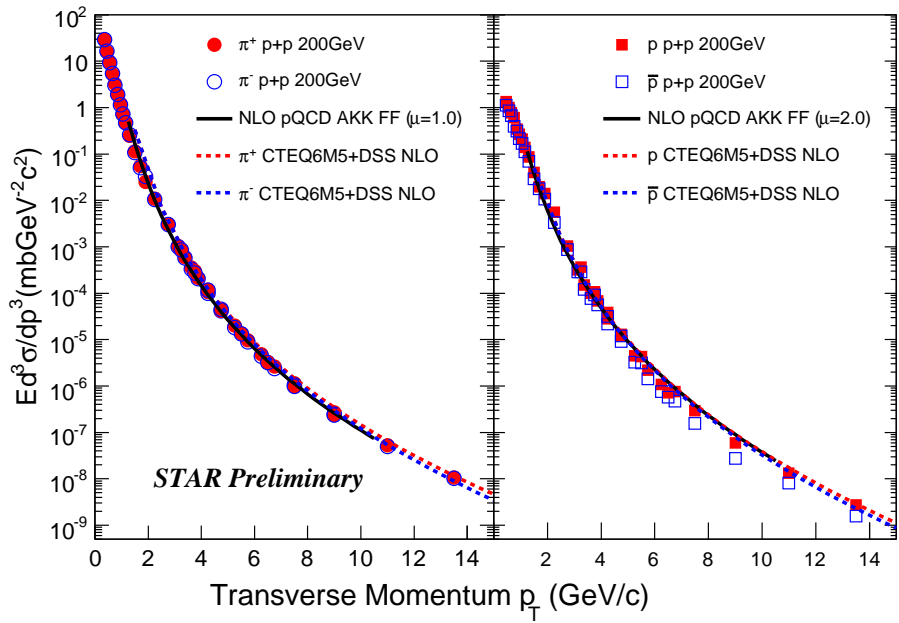}
\caption{Identified charged hadron spectra ($\pi^{\pm},p,\bar{p}$) in p+p collisions at $\sqrt{s}=200$ GeV. The curves are pQCD calculations with different fragmentation functions for pion and proton.}
\label{ppspectra}}
\end{figure}
In this talk, I have presented the recent development in constraints of fragmentation functions by measurements of identified hadrons in p+p collisions at RHIC, and new theoretical development of fragmentation functions with global fit to data provided by both $e^+e^-$ and hadron collisions~\cite{AKK,KKP,DSS,AKK2}. 
The single inclusive hadron spectra are limited to be {\bf $p_T{}^{>}_{\sim}5$} GeV/c to avoid complication of quark coalescence effect in A+A collisions and to provide a more reliable test bench for pQCD. I would then use this information to guide us in interpreting nuclear modification functions $R_{AA}$ of leading identified hadrons in central Au+Au collisions at RHIC. 

\section{Fragmentation Functions and Jet Chemistry in p+p collisions at RHIC}
The most uncertain part of the three terms relevant to RHIC physics is fragmentation function. In the $5<p_T<20$ GeV/c range at RHIC energy, most of the pions come from $qg\rightarrow qg$ and $qq\rightarrow qq$ processes (PYTHIA). Different pion fragmentation functions provided by DSS, AKK and KKP (abbreviations of the authors's last names) parameterizations can satisfactorily describe $\pi^{0}$ and $\pi^{\pm}$~\cite{AKK,AKK2,KKP,DSS}. Although baryon production may be more difficult to interpret/implement in QCD models (e.g. popcorn mechanism in Lund Model), it is no different from mesons in terms of parameterization of fragmentation functions provided enough data points with good accuracy in the relevant region.  However, fragmentation functions from quark and gluon to leading baryons (proton, $\Lambda$, etc.) are poorly constrained. This is especially true at high-z (fraction of leading hadron to the jet energy), where it is most relevant in jet quenching at RHIC. In Ref.~\cite{DSS}, the authors performed a global fit of proton and anti-proton fragmentation functions with data from SLAC, LEP and STAR data~\cite{starPIDpapers}. 
They concluded that "at the presently accessible range of transverse momenta ($p_T<7$ GeV/c) and at mid-rapidities the production of single-inclusive hadrons is mainly driven by gluon-induced processes and fragmentation, turning these data into the best constraint on the gluon fragmentation function $D^{p}_{g}$ at large value of $z$ currently available". The authors stated that, with extended measurements in $p_T$ to where the quark fragmentation becomes significant, these data will allow to separate quark-to-proton and anti-quark-to-proton fragmentation functions in the global fit. 

\begin{figure}[h]
\includegraphics[scale=0.6]{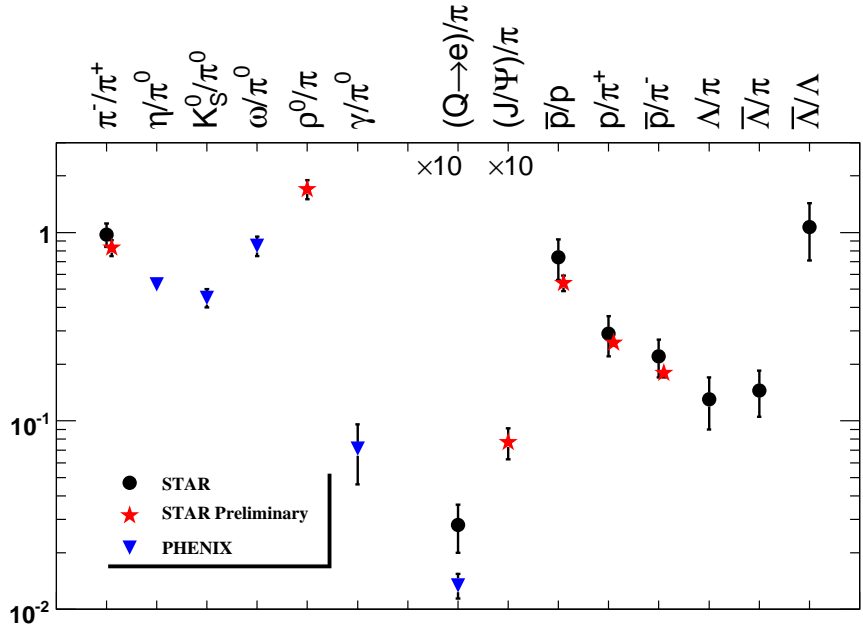}
\caption{The jet chemistry of Particle ratios at high $p_T$ in p+p collisions at $\sqrt{s}=200$ GeV. Particle ratios are obtained from particle yields with $p_T>5$ GeV/c except $\Lambda(\bar{\Lambda})/\pi$.}
\label{jetchemistry}
\end{figure}

PHENIX is able to reconstruct in the EMC mesons which decay to final-state photons and in the future with TOF and Aerogel Cherenkov Detector~\cite{phenixPIDQM06} for charged hadron identification. STAR/RHIC extends the particle identification of charged hadrons in Time Projection Chamber (TPC) from $p_T{}^{<}_{\sim}7$ GeV/c to $p_T{}^{<}_{\sim}15$ GeV/c. There are several improvements over the years in STAR, which make this extension possible: 
\begin{itemize}
\item Momentum and distortion calibration of TPC\cite{GeneTPC} 
\item Ionalization energy loss (dE/dx) calibration of TPC~\cite{PIDtech}
\item Jet-triggered data by EMC to increase the statistics at high $p_T$~\cite{starppjet} 
\end{itemize} 
\begin{figure}
{\includegraphics[scale=0.5]{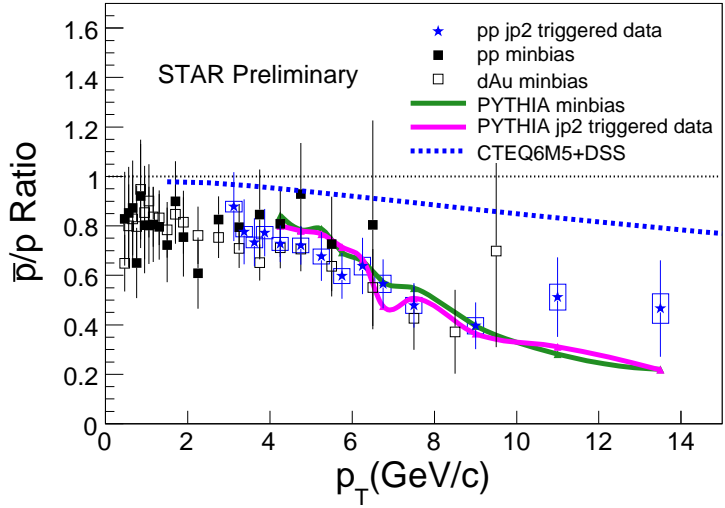}
\caption{$\bar{p}/p$ ratio as function of $p_T$ in p+p collisions at $\sqrt{s}=200$ GeV. In comparison is the ratio from pQCD+DSS and PYTHIA. Noted that other fragmentation models (KKP, AKK, Kretzer) did not distinguish particle and anti-particle and therefore the ratio is unity.}
\label{pbarpratio}}
\end{figure}
Fig.~\ref{ppspectra} shows the $\pi^{\pm}$, p and $\bar{p}$ spectra from minbias and jet-triggered p+p collisions. The results from jet-triggered event sample have been corrected for trigger bias~\cite{QMproceedings}, and are consistent with results from minbias data where two overlap. Most of the models (DSS, KKP and AKK) can reproduce the charged pion spectra quite well. We see slightly decrease of $\pi^{-}/\pi^{+}$ from unity at $p_T{}^{<}_{\sim}5$ GeV/c to 0.8 at $p_T\simeq15$ GeV/c. The difference is due to the contribution of valence-quark fragmentation. In the current KKP, AKK and Kretzler models, there is no distinction between $u\rightarrow\pi^{+}$ and $u\rightarrow\pi^{-}$. Neither is there difference between $u,d\rightarrow p$ and $u,d\rightarrow\bar{p}$. This means that $\pi^{-}/\pi^{+}=\bar{p}/p=1$. The DSS model includes this difference, and is found to describe the $\pi^{-}/\pi^{+}$ ratio as a function of $p_T$ as shown in Fig.~\ref{pbarpratio}. AKK and DSS can describe the proton yields reasonably well (within 20\%). However, DSS model tends to overpredict the ratio of $\bar{p}/p$ at high $p_T$ while AKK doesn't distinguish proton and anti-proton. The overprediction of $\bar{p}$ yields at high $p_T$ from DSS means that the fragmentation of quark-to-antiproton was over-estimated. The new AKK~\cite{AKK2} fragmentation functions implement the flavor dependence of parton fragmentation and predict a 10--20\% percent difference of $\pi^{-}$ to $\pi^{+}$ at $p_T=10$ GeV/c due to valence quark contribution. These ratios should provide a stringent constraint on quark and gluon contributions to identified hadrons at this $p_T$ range. Fig.~\ref{jetchemistry} shows the measured particle ratio in p+p collisions by STAR and PHENIX collaborations
~\cite{QMproceedings,phenixeta,starcharm,starstrange,phenixpi0}. 

\section{Effects of Jet Quenching on Jet Chemistry}
The light mesons are: $\pi^{\pm,0}$, 
$\eta$, $K^{0}_{S}$, $\rho^{0}$ and $\omega$; baryons are: $\Lambda(\bar{\Lambda})$, $p(\bar{p})$; in addition, heavy-flavor measurements are: $(Q\rightarrow e)$ and $J/\Psi$. The jet chemistry from particle ratios is very different from particle ratios from the total integrated yields at low $p_T$~\cite{starwhitepaper}. For example, pions are the dominant source of soft hadrons (bulk) while $\rho$ yields at high $p_T$ are high than pion yields~\cite{starrho}. The change of chemistry in the bulk provides a chemometer for assessing the chemical potential, temperature and strangeness equilibrium. Jet quenching changes the parton composition and fragmentation, resulting in possible change of hadron chemistry in out-going jets. Nuclear modification factor $R_{AA}$ is used to quantify the suppression of hadron yields in Au+Au collisions relative to $N_{bin}$-scaled yields in p+p collisions. We observed that light mesons ($\pi,\eta,\rho,K^{0}_{S}$) at high $p_T$ and electrons from heavy-flavor semileptonic decay have similar $R_{AA}$ while $p(\bar{p})$ $R_{AA}$ are systematically above the pion $R_{AA}$.  

\begin{figure}[h]
\includegraphics[scale=0.6]{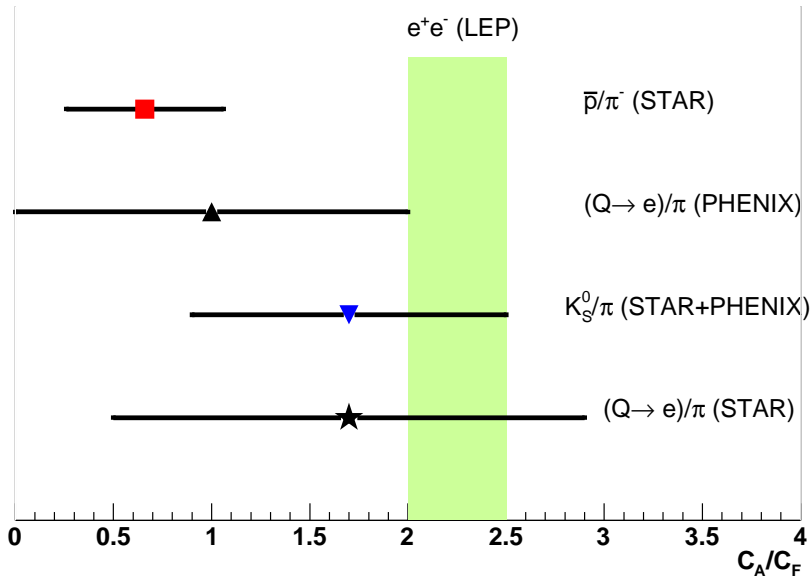}
\caption{Effective color-charge factor from $R_{AA}$ due to energy loss between gluon and quark energetic parton jets. The algebra is detailed in the text. This shows that the difference of energy loss between different partons is small, in contrast to the expectation of $C_A/C_F=9/4$ and dead cone effect. The band in the figure is the measurement of $C_A/C_F$ from $e^+e^-$ at LEP.}
\label{colorvalue}
\end{figure}

In AKK model and PYTHIA simulation as well, 60\% of $\pi^{\pm}$ are from quark fragmentation and 40\% from gluon fragmentation at $p_T=10$ GeV/c while that partition for ($p+\bar{p}$) is 10\% and 90\%, and that for $K^{\pm},K^{0}_{S}$ is 20\% and 80\%. This provides a tool for studying quark and gluon color-charge factor in jet quenching. In WHDG model~\cite{WHDG}, the $R_{AA}$ charm quark resulting from the radiative energy loss is very similar to that of the light quarks in the $p_T$ range accessible to us. If we separate the $R_{AA}$ of hadrons at parton sources, 
$$R_{AA}^{\pi}=0.6R_{AA}^{q}+0.4R_{AA}^{g},
R_{AA}^{p}=0.1R_{AA}^{q}+0.9R_{AA}^{g},
R_{AA}^{K}=0.2R_{AA}^{q}+0.8R_{AA}^{g},
R_{AA}^{c\rightarrow e}=R_{AA}^{Q}.$$
Since $R_{AA}^{q}\simeq C\times R_{AA}^{g}$~\cite{WHDG} where $C$ is the effective color-charge factor, which is 9/4 in pQCD model~\cite{xnwang}. 
If double ratio of hadron $R_{AA}$ is taken, this results in: 
$$R_{AA}(p/\pi)={{0.1C+0.9}\over{0.6C+0.4}},
R_{AA}(K/\pi)={{0.2C+0.8}\over{0.6C+0.4}},
R_{AA}(Q\rightarrow e/\pi)={{C}\over{0.6C+0.4}}.$$
In the analog to the measurement of $C_A/C_F$ in $e^{+}e^{-}$ collisions, we plot in Fig.~\ref{colorvalue} the effective color-charge factor ($C$) extracted from this very simple approach. One can see that the value is systematically lower than 9/4. When the WHDG model~\cite{WHDG} was used to fit the $R_{AA}$ of $\pi^{0}$ in $5<p_T<20$ GeV/c~\cite{phenixqhat}, it is evident that there is a stronger increase of $R_{AA}$ as function of $p_T$ in the model than that exists in the data. The $R_{AA}^{\pi^{0}}$ data points essentially show no $p_T$ dependence. At parton level, the $R_{AA}$ of light quarks and gluons in WHDG model show very little $p_T$ dependence. Since $R_{AA}^{q}>R_{AA}^{g}$ and quark contribution to final-state pion increases with $p_T$, the resulting $R_{AA}^{\pi}$ increases with $p_T$. On the other hand, if the energy loss were not sensitive to the different color-charge factor of quark and gluon ($R_{AA}^{q}=R_{AA}^{g}$) in the model, the resulting $R_{AA}^{\pi}$ would have shown little $p_T$ dependence. This seems to be consistent with the jet chemistry analysis of $R_{AA}$ among the identified hadrons, where the effective color-charge factor $C\simeq1$. 

\section{Discussions}\label{discussion}
What does it mean when $C\neq9/4$? As mentioned earlier, there are several effects which can destroy the proportionality of energy loss and color-charge factor. We attempt to select only the particles with $p_T>5$ GeV/c to avoid the potential contributions from quark coalescence at hadronization to the jet chemistry. It is not clear whether that is a sufficient high $p_T$. The color-charge factor was measured and compared to leading order (LO) and NLO at small $\alpha_{S}$ where pQCD is applicable. In a strongly interaction QGP, although the out-going jet energy is comparable to jets produced in 4-jet events at LEP energy in $e^+e^-$ collisions, the radiation or collision happens in a strongly interacting medium and with much smaller $q^{2}$. 

Ref.~\cite{liuconversion,liuFries} show that higher order contribution can change the in-coming leading quark into an out-going leading gluon. This effectively decreases the value of $C$. With a larger interaction cross-section than expected from pQCD, the conversion can describe the $\bar{p}/\pi^{-}$ ratio from decrease from p+p to Au+Au, to a reverse trend. The authors~\cite{liuFries} further suggested a test using $K/\pi$ ratio. The current $K^{0}_{S}/\pi$ measurement (p+p data from PHENIX and Au+Au from STAR) is not yet able to distinguish the models. 

Recently, there was a proposal~\cite{brodskypaper} of "color transparency and direct hadron production" to explain the baryon/meson production at intermediate $p_T$. I thought that this is another fancy term for quark coalescence. On the other hand, there may be a significant difference between this mechanism and quark coalescence at higher $p_T(>5$ GeV/c) and with vector mesons ($\phi,J/\Psi$) as presented by ~\cite{QMproceedings} where the effect of quark coalescence is small but the direct hadron production should be significant from this prediction. The proposed mechanism seems to be consistent with the measurements as noted in the Ref.~\cite{brodskypaper}. Another possibility is that the ridge, which carries the chemistry of the bulk with high baryon/meson ratio~\cite{QMproceedings}, extends to high $p_T$ and contaminates the jet spectra. Extending the study of ridge chemistry at lower $p_T$ ($<5$ GeV/c) to higher $p_T$ will be crucial to see how much the ridge contributes to the baryon yields at high $p_T$. We noted that there are many examples of effective field theory in QED condensed matter physics when a strong field/interaction exist in the medium. For example, the perfect 2D electron liquid in Fractional Quantum Hall Effect in QED has Chern-Simon theory. We hope that a better measurement of effective color-charge factor may provide us with important information about the effective interactions in QGP. 

\section{Summary}
In summary, RHIC has provided the first measurements of jet chemistry in both p+p collisions and Au+Au collisions at high $p_T$. Comparison of $R_{AA}$ among different particles shows that light mesons have similar $R_{AA}$ as expected from pQCD and quark/gluon fragmentation while baryons have similar (slightly larger but within systematical errors) $R_{AA}$ as light mesons unexpected from jet quenching and quark/gluon fragmentation. The gluon jets are not more suppressed than light-quark jets or heavy-quark jets. Therefore, the effective color-charge factor ($C_A/C_F$) is consistent with unity.   

\section{Acknowledgement}
The author would like to thank the organizer for invitation, and thank Adam Kocoloski, Yichun Xu, Zebo Tang, Lijuan Ruan, Patricia Fachini, Anne Sickles, Bedanga Mohanty, Simon Albino, Xin-nian Wang, Simon Wicks, Thorsten Renk and Werner Vogelsang for valuable discussions and providing the data points and pQCD curves. This work is supported in part by the 2004 Presidential Early Career Award (PECASE) and DOE.

\end{document}